# Trajectory control in idealized four-wave mixing processes in optical fiber

**Anastasiia Sheveleva[1], Pierre Colman[1], John. M. Dudley[2], Christophe Finot[1],***

[1] *Laboratoire Interdisciplinaire Carnot de Bourgogne, UMR 6303 CNRS-Université de Bourgogne-Franche-Comté, 9 avenue Alain Savary, BP 47870, 21078 Dijon Cedex, France*

[2] *Institut FEMTO-ST, Université Bourgogne Franche-Comté CNRS UMR 6174, Besançon, 25000, France.*

\* <u>*Corresponding author*</u>:
E-mail address: <christophe.finot@u-bourgogne.fr>

**Abstract:** The four-wave mixing process is a fundamental nonlinear interaction in Kerr media that can be described by a closed trajectory in the associated phase plane. We show here that it is possible to manipulate these trajectories and to connect two points that are not part of the same orbit. Our approach is based on a localized abrupt modification of the average power of the system. This mechanism is confirmed using different experimental realizations where iterative propagation in a short fiber segments mimics propagation in an idealized optical fiber.

## 1. Introduction

Thanks to their very low losses and their long interaction length, optical fibers have proven to be an essential tool to explore the very rich nonlinear dynamics resulting from the complex combination of dispersion with nonlinearity. The nonlinear Schrödinger equation (NLSE) is then a model of choice. Its best-known solution is the optical bright soliton [1] which we celebrate in this special issue, and which is able to propagate without altering its temporal and spectral intensity profiles. But other temporal structures also exist, such as solitons on finite background, where the interaction with a continuous background leads to a periodic temporal or spatial localization [2], including the particular case of the Peregrine soliton which exhibits double localization [3]. Other coherent solutions on a periodically modulated background also exist such as cnoidal or dnoidal waves [4].

Frequency-domain analysis of the NLSE sheds light on the spectral properties of these coherent waves through the four-wave mixing process, which allows energy exchange between different evolving spectral components. In the regime of focusing nonlinearity, this process leads to modulation instability that ultimately leads to the generation of high-repetition trains of ultrashort structures [5, 6]. In its simplest configuration, the degenerate four-wave mixing scenario consists of a continuous pump of high intensity that interacts with two other continuous components located symmetrically on both sides of this pump. In this context, the dynamics can be reduced to a system of three coupled differential equations whose behavior can be easily interpreted in the phase plane through two canonical conjugate variables [7]. The evolution of the system then follows closed orbits so that, for a given power of the system, the trajectories will never pass through a point located outside its orbit.

In this contribution, we seek to remove this fundamental limitation. We introduce a simple and general approach based on a discrete change in one of the system properties, namely the average power. Abruptly changing the propagation conditions allows us to switch from one orbit to another, and thus to connect any two points of the phase portrait. To experimentally verify this phase-space manipulation, we implement a specific setup as introduced in reference [8], where iterated programmed initial conditions are sequentially reinjected into an optical fiber, allowing experimental observation of the full dynamical phase-space topology. Three experimental examples are provided and validate the ability to efficiently connect (using a discrete power change) two states that do not belong to the same orbit, even if they are on opposite sides of the system separatrix boundary.

## 2. Theoretical background and principle of our approach

We first review the theoretical description of ideal FWM dynamics. In single mode fiber, the evolution of the slowly-varying electric field envelope $\psi(z,t)$ is governed by the NLSE [9]:

$$i\frac{\partial \psi}{\partial z} - \frac{\beta_2}{2}\frac{\partial^2 \psi}{\partial t^2} + \gamma |\psi|^2 \psi = 0, \tag{1}$$

with $z$ being the propagation distance and $t$ the time in a reference frame traveling at the group velocity. The group-velocity dispersion is $\beta_2$ and the nonlinear Kerr coefficient is $\gamma$. We consider wave mixing associated with the injection of a modulated pump wave $\psi_0$ with two sidebands at angular frequencies $\pm\omega_m$:

$$\begin{aligned}\psi(z,t) = \psi_0(z) + \psi_{-1}(z)\exp(i\,\omega_m\,t) \\ + \psi_1(z)\exp(-i\,\omega_m\,t)\end{aligned}, \tag{2}$$

In general, the injection of such a modulated signal in a fiber leads to the generation of multiple additional sidebands [10, 11]. However, when those higher-order sidebands can be neglected as in our experiments [8], the nonlinear dynamics of the pump and two sidebands can be described by only three coupled equations [7]:

$$\begin{cases} -i\dfrac{d\psi_0}{dz} = \gamma\left(|\psi_0|^2 + 2|\psi_{-1}|^2 + 2|\psi_1|^2\right)\psi_0 \\ \qquad\qquad + 2\gamma\,\psi_{-1}\,\psi_1\,\psi_0^* \\ -i\dfrac{d\psi_{-1}}{dz} + \dfrac{1}{2}\omega_m^2|\beta_2|\psi_{-1} = \gamma\left(|\psi_{-1}|^2 + 2|\psi_0|^2 + 2|\psi_1|^2\right)\psi_{-1} \\ \qquad\qquad + \gamma\,\psi_1^*\,\psi_0^2 \\ -i\dfrac{d\psi_1}{dz} + \dfrac{1}{2}\omega_m^2|\beta_2|\psi_1 = \gamma\left(|\psi_1|^2 + 2|\psi_0|^2 + 2|\psi_{-1}|^2\right)\psi_1 \\ \qquad\qquad + \gamma\,\psi_{-1}^*\,\psi_0^2 \end{cases}. \tag{3}$$

This system is equivalent to a one-dimensional conservative nonlinear oscillator with the Hamiltonian expressed as:

$$H = 2(1-\eta)\cos(\phi) - (\kappa-1)\eta - \frac{3}{2}\eta^2, \tag{4}$$

where $\kappa$ is defined as

$$\kappa = \frac{\beta_2}{\gamma} \frac{\omega_m^2}{P}, \tag{5}$$

and $P = |\psi_0|^2 + |\psi_{-1}|^2 + |\psi_1|^2$ is the total average power. Here, the canonical variables $\eta$ and $\phi$ are built from the transformation of the amplitudes $\psi_k(z)$ and phases $\varphi_k(z)$ of the evolving sidebands ($k = 0, \pm 1$) to $\eta$ and $\phi$ by:

$$\begin{cases} \eta = |\psi_0|^2 / P \\ \phi = \varphi_1 + \varphi_{-1} - 2\varphi_0 \end{cases}. \tag{6}$$

$\eta$ and $\phi$ are interpreted as the fraction of the total power in the central frequency component and the phase difference between the sidebands and the pump, respectively.

The dynamics on the ($\eta$ ; $\phi$) plane fully captures all the physics of this system [7, 12], including multiple Fermi-Pasta-Ulam-Tsingou recurrence cycles [13, 14], the existence of a separatrix [13] and stationary wave existence. Figure 1(a) illustrates the trajectories that are followed when starting from two points : ($\eta = 0.98$ ; $\phi = 0$, dotted lines) and ($\eta = 0.98$ ; $\phi = \pi$, dashed lines). The fiber parameters used correspond to experiment that use an anomalously dispersion fiber with $\beta_2 = -7.6 \cdot 10^{-3}$ ps$^2$/m and $\gamma = 1.7 \cdot 10^{-3}$/W/m. We plot the orbits for three values of $P$ : 22, 23 and 24 dBm (green, purple and red lines, respectively) that exceed the power (21.7 dBm) associated with the maximal small-signal gain at the modulation frequency $\omega_m/2\pi = 40$ GHz. The trajectories are closed orbits that, for a given average power, never cross. The orbits obtained for $\phi = 0$ are localized on the right-hand side of the ideal FWM separatrix orbit and are followed counterclockwise, whereas orbits obtained for $\phi = \pi$ are on the left side and are followed clockwise. For our values of $\eta$ very close to unity, when they approach the central part of the phase plane, both orbits are in the vicinity of the separatrix.

If there exists no limitation in possible values of $\kappa$, any two points on the phase diagram can be in principle connected. Indeed, if two points are part of the same orbit, then, according to the four-wave ideal model, they share the same values of $\kappa$ and $H$. It is therefore possible to derive a value of $\kappa$ would connect two arbitrary states ($\eta_1$, $\phi_1$) and ($\eta_2$, $\phi_2$). Indeed, assuming the conservation of the Hamiltonian $H(\eta_1, \phi_1) = H(\eta_2, \phi_2)$, Eq. (4) leads to :

$$\begin{aligned}\kappa = \ & 2\left[1-(\eta_1+\eta_2)\right]\cos\left(\frac{\phi_1+\phi_2}{2}\right)\cos\left(\frac{\phi_2-\phi_1}{2}\right)\\ & +1-\frac{3}{2}(\eta_1+\eta_2)\\ & -2\sin\left(\frac{\phi_1+\phi_2}{2}\right)\sin\left(\frac{\phi_2-\phi_1}{2}\right)\frac{\eta_1+\eta_2-\eta_1^2-\eta_2^2}{\eta_2-\eta_1}\end{aligned} \qquad (7)$$

Except for the case $\eta_1=\eta_2$, where no solution exists except for the trivial case where $\phi_1=\pm\phi_2$, we see that two arbitrary points can indeed be connected. However, one has to take into account several strong restrictions regarding the accessible values of $\kappa$ : the dispersion regime limits, for a given fiber, $\kappa$ to positive or negative values, while the ratio $\beta_2/\gamma$ is not fully flexible either, and moreover the total power $\gamma P$ has to be limited so as to remain in the framework of Eq. (3). Therefore the approach based on Eq. (7) only cannot in practice be applied in a straightforward manner. Consequently, when $\kappa_{min}<\kappa<\kappa_{max}$, a different strategy has to be developed. One example of the typical problem we want to solve is to find a way to propagate from an input ($\eta_{IN}$, $\phi_{IN}$) to output parameters ($\eta_{OUT}$, $\phi_{OUT}$) that cannot be located on the same orbit for the range of $\kappa$ under investigation. Figure 1(a) illustrates this problem with the input ($\eta_{IN}$, $\phi_{IN}$) being (0.98 , $\pi$) (red diamond) and the output ($\eta_{OUT}$, $\phi_{OUT}$) being (0.98, 0) (cyan diamond). The input and output being located on different sides of the separatrix, it is in principle impossible to connect them. In order to overcome this fundamental limit, we must imprint a change in the physical parameters of the system. Such an approach has been investigated for example to freeze the evolution of a breathing state evolution in split-dispersion cascaded photonic crystal fibers [15] or more recently in the context of control of oceanic waves that are also described by the NLSE equation [16]. In both cases, the change was affecting the dispersive or nonlinear properties of the propagation medium with for example a change in the optical fiber design [15] or an abrupt change in the depth of the water flume [16], enabling the conversion of an Akhmediev breather into a dnoidal wave. More generally speaking, it has been theoretically discussed that the parameter $\kappa$ is the crucial control parameter to act on the breathing state properties and reach a stationary point [17].

In this contribution, we explore a more general approach where we do not limit our analysis to a target state being a stationary point. We want to connect two arbitrary states of the phase plane. In order to modify the control parameter $\kappa$, we physically rely on an abrupt change of the average power $P$. Indeed, the average power strongly influences the trajectory as well as the location of the separatrix, as experimentally demonstrated in [8]. $\kappa_{min}$ and $\kappa_{max}$ will in this case be defined by the maximal and minimal values of the power $P_{max}$ and $P_{min}$ that are available. A single change is required when the switching point is chosen with care. Indeed, as shown in Fig. 1(a), for a good combination of input and output powers $P_{IN}$ and $P_{OUT}$, crossing between the two orbits (including in one case ($\eta_{IN}$, $\phi_{IN}$) for power $P_{IN}$ and in the other case ($\eta_{OUT}$, $\phi_{OUT}$) for power $P_{OUT}$) may occur. The crossing points are marked with black dots in Fig. 1(a) where we can make out that with a combination of only 2 powers chosen among 3 discrete possibilities, 6 crossing points may exist.

As the power can be selected on a continuous range, the number of possible combinations is in fact much higher. In order to find more systematically all the switching points ($\eta_S$ ; $\phi_S$) that are possible for $\kappa_{min} < \kappa < \kappa_{max}$, we scan the full space ($\eta_{TMP}$ ; $\phi_{TMP}$). Using Eq. (7), we determine the values $\kappa_{IN}$ out $\kappa_{OUT}$ required to connect ($\eta_{IN}$, $\phi_{IN}$) to ($\eta_{TMP}$ ; $\phi_{TMP}$) and ($\eta_{TMP}$ ; $\phi_{TMP}$) to ($\eta_{OUT}$, $\phi_{OUT}$), respectively. If both $\kappa_{IN}$ out $\kappa_{OUT}$ are confined within [$\kappa_{min}$ ; $\kappa_{max}$] then ($\eta_{TMP}$ ; $\phi_{TMP}$) represents a possible switching point. This leads to the colored area in Fig. 1 (b)) obtained for $\kappa \in$ [-1.9 ; -1.2] (corresponding to powers between 22 and 24 dBm).

The question that then arises is the choice of the optimal switching point. Here, we define this optimal solution as the solution leading to the shortest propagation distance, but other criteria can be chosen. To answer this, we performed numerical simulations of the system to evaluate the physical distance of propagation to link ($\eta_{IN}$, $\phi_{IN}$) to ($\eta_{OUT}$, $\phi_{OUT}$) with a single power change in between. Results are plotted with a colormap in Fig. 1(b) that stresses that, according to the choice of $P_{IN}$ and $P_{OUT}$, the distance to be involved may vary by 24 % from 16.19 km to 20.65 km. The optimum combination is found to be $P_{IN}$ = 23.47 dBm ($\kappa_{IN}$= -1.35) and $P_{OUT}$ = 23.99 dBm ($\kappa_{OUT}$= -1.20) leading to the trajectory plotted in red (when the power is $P_{IN}$) and in cyan (when the power is $P_{OUT}$). The location ($\eta_S$ ; $\phi_S$) at which the power switching occurs is highlighted on the phase plane by a black circle and can also be seen in Fig. 1(c) where the longitudinal evolution of the average power is displayed.

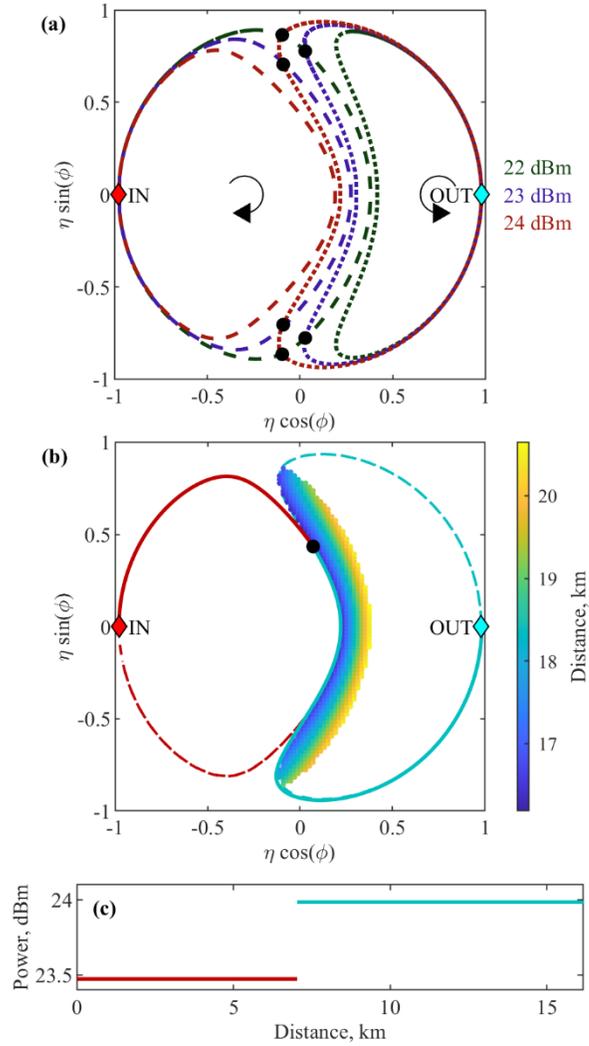

**Figure 1. (a)** Phase space portraits. The orbits including the states ($\eta_{IN}=0.98; \phi_{IN}=\pi$) and ($\eta_{OUT}=0.98; \phi_{OUT}=0$) are plotted with dashed and dotted lines respectively. Results for average powers of 22, 23 and 24 dBm are plotted with green, purple and red colors respectively. The black circles highlight the crossing points of two trajectories. **(b)** Optimal trajectory followed from ($\eta_{IN}; \phi_{IN}$) to ($\eta_{OUT}; \phi_{OUT}$) (solid line). The orbits including the input and output states are plotted with dashed lines. The part corresponding to the average power $P_{IN}$ and $P_{OUT}$ is plotted with red and cyan colors respectively and the switching point is marked with a black circle. The area corresponding to possible switching locations according to the combination of input and output powers is colored with a colormap linked to the distance needed to link the test states. **(c)** Longitudinal evolution of the average power used for the optimum solution.

## 3. Experimental setup

Experimentally recording the longitudinal evolution of the optical field remains a hard task. In order to avoid destructive approaches such as cut-back measurements [18] or the involvement of multiple fiber segments [19], various techniques have been tested: advanced shaping of combs [10], distributed optical time domain reflectometry [20, 21] or evolution in a recirculating loop [22, 23]. However, generation of new unwanted spectral lines resulting from the cascading of the four-wave mixing process remains an issue, as well as the impact of losses that have to be managed with extreme care. By using iterated programmed initial conditions that are sequentially reinjected into the fiber under investigation, the experimental approach we implemented relaxes those constraints. The setup is identical to the one detailed in [8] and is shown in Fig. 2. It relies on optical components from the telecommunication industry. A continuous wave (CW) laser operating at 1550 nm is first sinusoidally modulated using a 40-GHz phase modulator (PM) to create an equispaced frequency comb. The RF amplitude of the modulation is chosen so as to approach a level of the three central optical components to be roughly equal. The resulting symmetrical comb is then processed using a programmable filter [24] to tailor the pump and the two coherent seeds with the target $\eta_i$ and $\phi_i$, and to filter out any unwanted higher order harmonics. The comb is then amplified by an erbium-doped fiber amplifier (EDFA) that delivers a constant average power $P$ that can be varied between 22 and 24 dBm ($\kappa$ between -1.9 and -1.2).

Propagation occurs in a 500 m length of fiber with the dispersive and nonlinear parameters as given in the previous section. The impact of losses over such a distance is negligible and does not require any compensation schemes. The length of the segment has been chosen as a tradeoff between the sensitivity of the detection of the changes of the wave properties (enough dispersion or nonlinearity should be accumulated) and the appearance of detrimental effects such as Brillouin or Raman scattering. We have also checked that, with half a kilometer propagation distance, the growth of additional sidebands remains low enough so that the framework of a degenerate four-wave interaction remains valid.

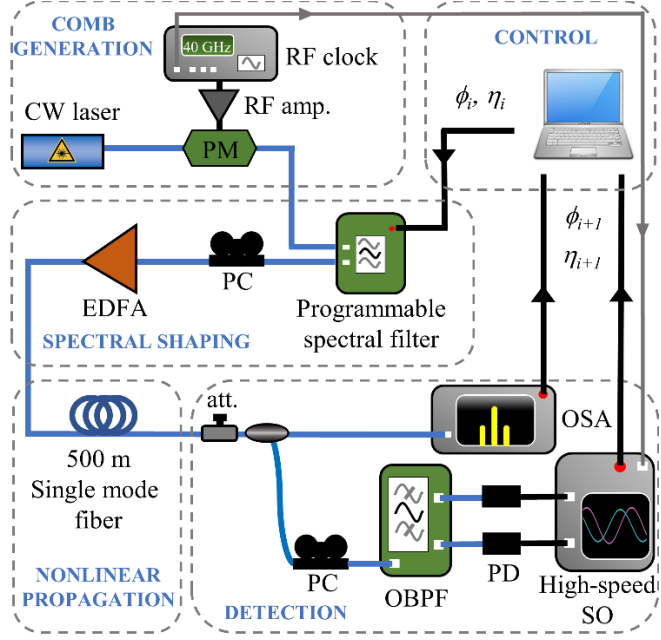

**Figure 2.** Experimental setup. PC: polarization controller. Att: Attenuator. OBPF: optical band pass filter. PD: photodiode. RF amp: Radiofrequency amplifier. Other abbreviations are described in the text.

The output signal is split into two channels in order to simultaneously record both spectral phase and amplitude. An optical spectrum analyzer (OSA, resolution 0.1 nm) provides directly the ratio $\eta_{i+1}$. The spectral phase offset $\phi_{i+1}$ is retrieved from the temporal delay between the central and lateral sidebands as measured by photodiodes connected to a high-speed sampling oscilloscope (bandwidth of the optoelectronic chain being larger above 40 GHz).

The process is then iterated and the experimentally measured values are imprinted as new inputs. Note that we recently proposed an alternate method benefiting from machine learning strategies based on a large set of initial random measurements that are then processed using artificial neural networks [25].

The transition between the two orbits connecting ($\eta_{IN}$ ; $\phi_{IN}$) to ($\eta_{OUT}$ ; $\phi_{OUT}$) is achieved using a change between two measurements in the average power of the wave, switching from $P_{IN}$ to $P_{OUT}$ after the prescribed propagation distance is attained. Using the power as a tuning variable in the control parameter $\kappa$ provides much more flexibility than abruptly changing the dispersive or nonlinear waveguide properties for which only a few values are available and that require a replacement of the fiber. Note that given the discreteness of the propagation distances under investigation (the distances being multiples of the 500m length of the fiber segment), we cannot carry the change in power at the exact distance predicted by the numerical analysis. We therefore monitor when ($\eta_{i+1}$ ; $\phi_{i+1}$) approaches ($\eta_S$ ; $\phi_S$) by checking the distance $D_{i+1}$ (when plotted on the phase plane) between the state $i+1$ and the switching point. When we operate in the vicinity of the switching coordinates and when $D_{i+1} > D_i$, we update the state back to ($\eta_i$ ; $\phi_i$) and carry out the power change. By doing so we take into account possible deviations from the theoretical model, and we approach the switching position at the nearest.

## 4. Experimental results for the trajectory manipulation

Let us first discuss the case where the input and output state lies on the same side of the separatrix. We have considered here two examples summarized in Fig. 3, with panels (1) and (2) corresponding to cases where the two states lie on the right and left sides of the separatrix, respectively. The input and output states have been chosen with the same phase ($\varphi = 0$ or $\pi$), but similar results could be obtained with different values. The target is to connect ($\eta_{1IN} = 0.90$, 0) to ($\eta_{1OUT} = 0.80$, 0) (case 1) and ($\eta_{2IN} = 0.90$, $\pi$) to ($\eta_{2OUT} = 0.80$, $\pi$) (case 2). Panels (a) are obtained from numerical simulations and show the area where the power switch should be achieved when the average power available ranges between 22 and 24 dBm. Note that the higher the range of power available is, the broader this area is. The optimum combination is marked with a black circle and corresponds to a power change from $P_{1IN} = 22.7$ dBm to $P_{1OUT} = 23.8$ dBm (case 1) and from $P_{2IN} = 23.5$ dBm to $P_{2OUT} = 22.9$ dBm (case 2). Whereas in the case 1, one has to increase the power in order to reach a point in an inner orbit, this is the contrary in the second case. The corresponding switching propagation distance predicted by numerical simulations occurs at 6.5 km and 11.5 km, for case 1 and 2 respectively.

The longitudinal variation of the average power that has been experimentally implemented is plotted on panels (b). The corresponding trajectories mapped on the phase-space portraits are reported on panel (c). Given the granularity of the distances involved in the experiment as well as some sources of errors in the measurements of the wave parameters, some deviations would appear with respect to the ideal orbits. However, those discrepancies remain very slight so that, for these two cases, we can convincingly validate the numerical predictions and conclude to the efficient control of the trajectory.

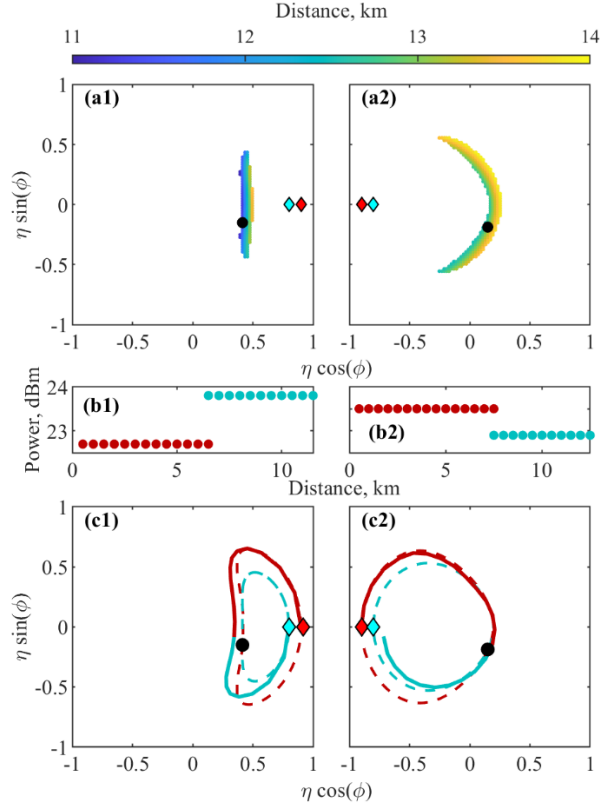

**Figure 3.** Illustration of the trajectory connecting two states located on the same side of the separatrix. Panels (1) deals with ($\eta_{1IN} = 0.90$, 0) to ($\eta_{1OUT}= 0.80$, 0) whereas panels two investigate and ($\eta_{2IN} = 0.90$, $\pi$) to ($\eta_{2OUT}= 0.80$, $\pi$) **(a)** Input states ($\eta_{IN}$; $\phi_{IN}$) and connected states ($\eta_{OUT}$; $\phi_{OUT}$) (red and cyan diamonds, respectively). The area corresponding to possible switching locations according to the combination of input and output powers is colored with a colormap linked to the distance between the two states estimated by numerical simulations. The optimal switching point is marked with a black circle. **(b)** Longitudinal evolution of the average power used for the optimum trajectory experimentally implemented. **(c)** Trajectory followed experimentally from ($\eta_{IN}$; $\phi_{IN}$) to ($\eta_{OUT}$; $\phi_{OUT}$) (solid line). The orbits including the input and output states are plotted with dashed lines. The part corresponding to the average power $P_{IN}$ and $P_{OUT}$ is plotted with red and cyan colors respectively and the switching point predicted by numerical simulations is marked with a black circle.

In a second set of experiments, we investigate the connection between two states that are located on both sides of the separatrix, i.e. ($\eta_{IN} = 0.90$ ; $\phi_{IN} = \pi$ ) to ($\eta_{OUT} = 0.90$ ; $\phi_{OUT} = 0$). Following an analysis similar to the one reported in the previous section, we found numerically the optimum switching point. The experimental validation is summarized in Fig. 4 and further validates our approach to this challenging target. Once again, a single change of average power by less than 2dB imprinted at a distance of 6.5 km was required to efficiently reach the output. As in the experiment reported in Fig. 3(c2), the slight mismatch between the experimental output state and the target is mainly due to the finite length of 500m of the segments of fibers (as the power involved here is quite significant, the discreteness of the states that are accessed becomes more visible).

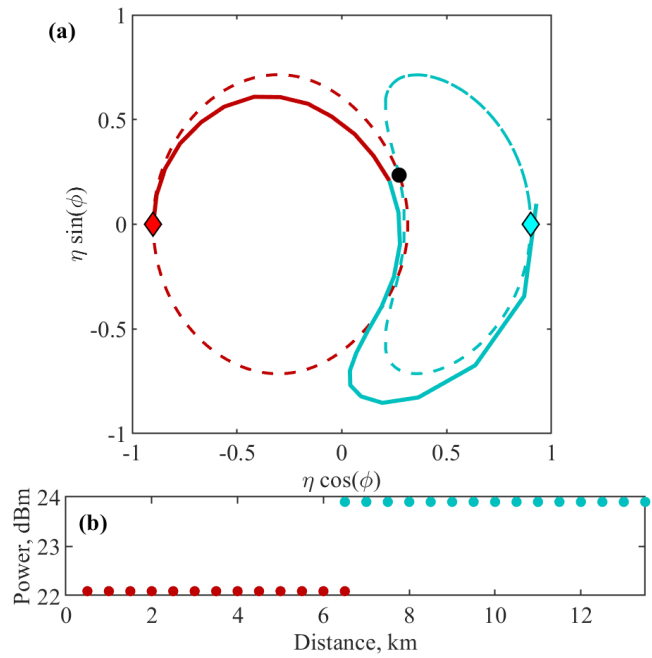

**Figure 4. (a)** Trajectory followed experimentally from ($\eta_{IN} = 0.90$ ; $\phi_{IN} = \pi$ ) to ($\eta_{OUT} = 0.90$; $\phi_{OUT} = 0$) (solid line). The orbits including the input and output states are plotted with dashed lines. The part corresponding to the average power $P_{IN}$ and $P_{OUT}$ is plotted with red and cyan colors respectively and the switching point predicted by numerical simulations is marked with a black circle. **(b)** Longitudinal evolution of the average power used for the optimum trajectory experimentally implemented.

## 5. Conclusions

To conclude, we have demonstrated here that the four-wave mixing interaction in optical fiber can be efficiently manipulated by using a control parameter such as the total power. A single local and abrupt change of the average power after a well-chosen propagation distance enables one to manipulate the trajectories and to jump from one orbit to another. This allows us to connect two states even if they lie on different sides of the separatrix. A simple and general methodology to select the suitable switching states has been presented, and experiments taking advantage of iterated programmed initial conditions confirm this approach. Three examples have been discussed, but we have also successfully tested many other combinations of input and output states. Whereas our demonstration is made in the framework of dynamics strictly ruled by a non-cascaded four-wave mixing, the proposed scheme could also be relevant to process NLSE-based dynamics.

Although we have focused our investigation here on the use of a single change of power in a focusing nonlinearity where the input and output average powers are the degree of freedom, this work can be further extended in many directions. First, one may choose to fix the input and output powers. In that case, if the input and output orbits do not cross, a third power level will have to be involved as an intermediate stage. Then, more advanced longitudinal variation profiles could be considered [26] with the goal for example being to decrease the distance required to perform the adiabatic transition between the two states. Such an analysis could also be of interest to better understand the impact of losses or distributed amplification such as Raman scattering [27]. In that context, our experimental setup is well suited to approach a continuous power profile with a set of 500m steps. Finally, dynamics involving 4 symmetrical spectral lines could be considered, the orbits being followed being more complex [28].

## Funding

## Acknowledgments

The authors also thank GDR Elios (GDR 2080) and very fruitful discussions with Bertrand Kibler.

## Declaration of Competing Interest.

The authors declare that they have no known competing financial interests or personal relationships that could have appeared to influence the work reported in this paper.

## CRediT authorship contribution statement

**Anastasiia Sheveleva:** Formal analysis, Methodology, Investigation, Software, Visualization, Writing - review & editing. **Pierre Colman:** Validation, Writing - review & editing. **John M. Dudley:** Validation, Writing - review & editing, Funding acquisition. **Christophe Finot**: Conceptualization, Methodology, Validation, Writing-original draft, Supervision, Funding acquisition.

## Data availability

The data that support the findings of this study are available from the corresponding author, CF, upon reasonable request.